# An Equivalent Volume Law for Anisotropic Laminated Structures

Mehmet Zor

Department of Mechanical Engineering, Dokuz Eylul University, Izmir, Turkiye, mehmet.zor@deu.edu.tr

**Abstract**

The problem of representing laminated structures by an equivalent volume and determining the elastic constants of this equivalent volume from the layer properties is a fundamental issue in the analysis of composite and multilayered systems. In the literature, the most widely used approach for this purpose is the Voigt-type volume-weighted averaging method. Although this method is widely accepted in practice, the uniqueness (as a material characteristic) of the equivalent elastic constants, their independence from the applied loading, and the mathematical conditions under which they can be defined have not been made explicit. This issue is particularly unclear for structures with asymmetric stacking sequences and general anisotropic behavior, including triclinic cases.

In this study, a theoretical framework referred to as the Zor model is presented, and it is shown that the elastic constants of the equivalent volume can be obtained only under the action of all in-plane force components ($F_x + F_y + F_{xy}$), directly from static equilibrium, linear elasticity, and perfect bonding conditions between the layers. No additional assumptions are introduced in the formulation; reciprocity is not assumed at the system level, but instead emerges naturally as a result of the solution process. The resulting equivalent elastic constants are independent of the applied loading and represent the intrinsic mechanical characteristics of the laminated structure.

## 1. INTRODUCTION

Laminated and multilayered structures are widely used in engineering applications due to their high stiffness-to-weight ratio and directional mechanical performance. To represent the mechanical behavior of such structures at the global level, it is essential to define an equivalent volume and to determine the equivalent elastic constants of this volume from the properties of the individual layers.

Among the various equivalent-volume approaches proposed in the literature, the Voigt method, in which the elastic constants of the layers are averaged according to their volume fractions, is the most widely used [1,2,3]. This approach has found broad application in engineering practice, including anisotropic cases. However, the Voigt model does not provide a rigorous theoretical proof explaining why and under what mathematical conditions the equivalent elastic constants are unique material characteristics, why they are independent of the applied loading, or why reciprocity should hold at the system level. These properties are usually introduced implicitly as assumptions.

This ambiguity becomes particularly significant for asymmetric stacking sequences and generally anisotropic structures (including triclinic behavior), where the directional elastic response and interlayer interactions are more complex. For such systems, it is not clear under which loading conditions the equivalent elastic constants can be uniquely defined and what mathematical necessity underlies their existence.

This study makes explicit the mathematical necessity underlying the equivalent relations that are implicitly used in the literature. Within the framework of the Zor model, it is shown that the elastic constants of the equivalent volume arise necessarily from static equilibrium, linear elasticity, and perfect bonding when all in-plane force components ($F_x + F_y + F_{xy}$) are applied. In this formulation, reciprocity is not imposed as an a priori condition at the system level, but instead emerges automatically from the mathematical structure of the solution.

In this way, the proposed approach does not define equivalent elastic constants as empirical or heuristic averages, but as load-independent and unique material characteristics of generally anisotropic laminated structures, derived from fundamental mechanical principles. Since general anisotropy (triclinic symmetry) represents the most general linear elastic material class, all other material types — including orthotropic and isotropic — are obtained as special cases of the present formulation.

**Materials and Methods**

**2.1 Hooke Relation for General Anisotropic Materials**

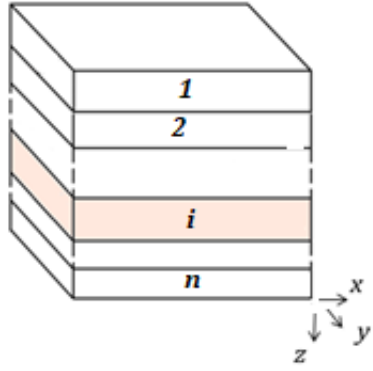

**Figure 1.** Laminated structure composed of anisotropic layers

For a layer *i*:

$$\begin{bmatrix} \sigma_{i_x} \\ \sigma_{i_y} \\ \tau_{i_{xy}} \end{bmatrix} = \begin{bmatrix} C_{i_{11}} & C_{i_{12}} & C_{i_{16}} \\ C_{i_{12}} & C_{i_{22}} & C_{i_{26}} \\ C_{i_{16}} & C_{i_{26}} & C_{i_{66}} \end{bmatrix} \begin{bmatrix} \varepsilon_{i_x} \\ \varepsilon_{i_y} \\ \gamma_{i_{xy}} \end{bmatrix} \quad (1)$$

Note that by taking $C_{i_{kj}} = C_{i_{jk}}$ on a layer-by-layer basis, we assume that condition Reciprocity condition is satisfied. Layer-level reciprocity is assumed as required by linear elasticity; system-level reciprocity is not assumed and is proved

For the entire layered structure:

$$\begin{bmatrix} \sigma_x \\ \sigma_y \\ \tau_{xy} \end{bmatrix} = \begin{bmatrix} C_{11} & C_{12} & C_{16} \\ C_{21} & C_{22} & C_{26} \\ C_{61} & C_{62} & C_{66} \end{bmatrix} \begin{bmatrix} \varepsilon_x \\ \varepsilon_y \\ \gamma_y \end{bmatrix} \quad (2)$$

We initially assume that $C_{kj} \neq C_{jk}$ for the entire layered structure. Our goal is to demonstrate that $C_{kj} = C_{jk}$, thus proving that Reciprocity is satisfied for the entire structure.

Under single loading cases (pure $F_x$, pure $F_y$, or pure $F_{xy}$), the strain components are not independent. Therefore, a polynomial identity cannot be established and the stiffness coefficients cannot be defined from these cases alone. The relations derived under single loadings should be interpreted only as projections of the equivalent law obtained under the general in-plane loading condition.

### 2.1 Under pure $F_x$ loading:

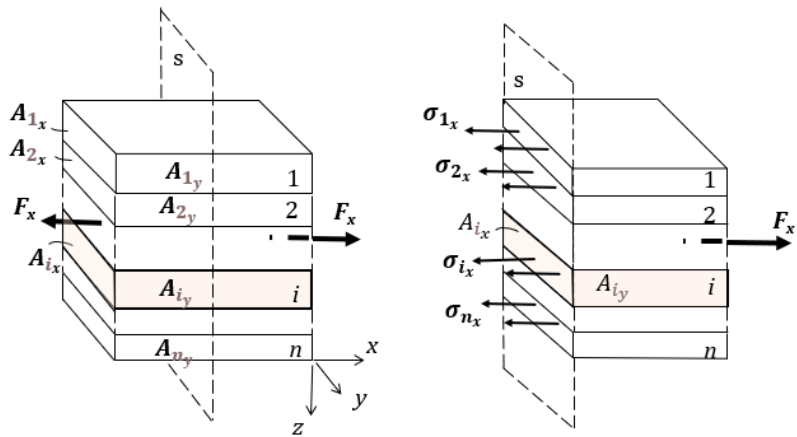

Figure 2

Only due to the $F_x$ force and perfect bonding,

the strains of all layers and the entire system in the x are equal: $\varepsilon_{i_x} = \varepsilon_x$ (3a)

Moreover, because a pure uniaxial loading $F_x$ does not produce any interlaminar shear deformation in a perfectly bonded laminate, the engineering shear strains must automatically vanish:

$$\gamma_{i_{xy}} = \gamma_{xy} = 0 \tag{3b}$$

**Poisson effect**: In addition to causing elongation in the x-direction, the applied tensile force $F_x$ also tends to contract the structure in the y-direction. Because of the

(3c)

perfect bonding between layers with different elastic properties and Poisson's ratios, transverse strain $\varepsilon_y$ becomes equal across all layers.

$$\varepsilon_{i_y} = \varepsilon_y$$

*Note: Because the layers possess different elastic properties and Poisson ratios, the perfect bonding condition forces them to undergo the same transverse strain $\varepsilon_y$. This mechanical constraint causes some layers to contract less than their natural Poisson tendency, while others are forced to contract more. Consequently, internal constraining forces $(F_{i_y})$ develop in the y-direction of each layer, accompanied by corresponding normal stresses $(\sigma_{i_y})$. However, when the external load acts only in the x-direction, the total transverse force on the entire structure must be zero $(F_y = 0)$, which implies that the equivalent transverse stress is also zero $(\sigma_y = 0)$. In a separate study that has been submitted to a peer-reviewed journal, this Poisson interaction between layers was explicitly incorporated into the formulation of an equivalent volume, and closed-form analytical expressions were obtained using only static equilibrium, Hooke's law, and full adhesion conditions*

The total externally applied axial force $F_x$ in the x-direction is equal to the sum of the internal axial forces carried by the layers:

$$F_x = F_{1_x} + F_{2_x} \dots F_{i_x} + \cdots F_{n_x} \rightarrow F_x = \sum_{i=1}^{n} F_{i_x} \tag{4}$$

These forces are equal to the product of the normal stresses and the corresponding cross-sectional areas. Therefore:

$$\sigma_x A_x = \sigma_{1_x} A_{1_x} + \sigma_{2_x} A_{2_x} \dots \sigma_{i_x} A_{i_x} + \cdots \sigma_{n_x} A_{n_x} \tag{5a}$$

Dividing both sides by $A_x$ we obtain:

$$\frac{\sigma_x A_x}{A_x} = \frac{\sigma_{1_x} A_{1_x}}{A_x} + \frac{\sigma_{2_x} A_{2_x}}{A_x} \dots \frac{\sigma_{i_x} A_{i_x}}{A_x} + \cdots \frac{\sigma_{n_x} A_{n_x}}{A_x} \tag{5b}$$

Thus, the normal stress in the equivalent volume in the x-direction is the weighted sum of the stresses in each layer (Eqs. 4 and 5, Figure 2).

(Volume fraction of the i-th layer: $V_i = \frac{A_{i_x}}{A_x}$ )

$$\sigma_x = \sigma_{1_x} V_1 + \sigma_{2_x} V_2 \dots \sigma_{i_x} V_i + \cdots \sigma_{n_x} V_n \tag{6}$$

More generally:

$$\sigma_x = \sum_{i=1}^{n} V_i \sigma_{i_x} \tag{7}$$

Layer-wise Hooke's law in the x-direction :

$$\sigma_{i_x} = C_{i_{11}} \varepsilon_x + C_{i_{12}} \varepsilon_y \tag{8}$$

$$\sigma_x = \sum_{i=1}^{n} V_i C_{i_{11}} \varepsilon_x + \sum_{i=1}^{n} V_i C_{i_{12}} \varepsilon_y \tag{9}$$

We can also write $\sigma_x$ from the equivalent stiffness matrix. So;

Equivalent expression in the x-direction:

$$\sigma_x = C_{11} \varepsilon_x + C_{12} \varepsilon_y \tag{10}$$

(Here $C_{11}$ and $C_{12}$ are the terms of the equivalent stiffness matrix).

Since equations (9) and (10) are equal to each other:

$$C_{11} \varepsilon_x + C_{12} \varepsilon_y = \sum_{i=1}^{n} V_i C_{i_{11}} \varepsilon_x + \sum_{i=1}^{n} V_i C_{i_{12}} \varepsilon_y \tag{11}$$

## 2.2 Under pure $F_y$ loading:

Similar procedures apply to the F$_y$ loading. Although $\sigma_y$ stress occurs, other stresses do not occur on a system basis ($\sigma_x = \tau_{xy} = 0$). Therefore, the following similar results are obtained.

Layer-wise Hooke's law in the y-direction:

$$\sigma_y = \sum_{i=1}^{n} V_i C_{i_{12}} \varepsilon_x + \sum_{i=1}^{n} V_i C_{i_{22}} \varepsilon_y \tag{12}$$

Equivalent expression in the y-direction:

$$\sigma_y = C_{21}\varepsilon_x + C_{22}\varepsilon_y \tag{13}$$

$$C_{21}\varepsilon_x + C_{22}\varepsilon_y = \sum_{i=1}^{n} V_i C_{i_{12}} \varepsilon_x + \sum_{i=1}^{n} V_i C_{i_{22}} \varepsilon_y \tag{14}$$

## 2.3 Under pure $F_{xy}$ loading:

When defining the shear modulus G$_{xy}$, the standard 'pure shear' strain path ($\varepsilon_x = \varepsilon_y = 0$) is adopted. This is not done to artificially eliminate the coupling terms $C_{16}$ and $C_{26}$, but is a kinematic requirement inherent in the definition of shear modulus; the equivalent C-matrix itself is determined independently from the general loading case.

The plane shear force $F_{xy}$ causes only shear stresses $\tau_{xy}$ to occur on the entire structure basis, while normal stresses do not occur ($\sigma_x = \sigma_y = 0$).

$$\tau_{xy} = \sum_{i=1}^{n} V_i \tau_{i_{xy}} = \sum_{i=1}^{n} V_i C_{i_{66}} \gamma_{i_{xy}} \tag{15}$$

$$\tau_{xy} = C_{66} \gamma_{xy} \tag{16}$$

$$C_{66} \gamma_{xy} = \sum_{i=1}^{n} V_i C_{i_{66}} \gamma_{i_{xy}} \tag{17}$$

Under single-component loading cases such as pure $F_x$, pure $F_y$, or pure $F_{xy}$, the strain components are no longer independent. The zero-force constraints impose relations among $\varepsilon_x$, $\varepsilon_y$ and $\gamma_{xy}$, so the strain space is not fully spanned. Consequently, the resulting stress–strain equations are valid only along a restricted loading path and do not constitute a polynomial identity. Therefore, the complete set of equivalent elastic constants cannot be uniquely determined from such single-component loadings.

In contrast, when all in-plane force components ($F_x + F_y + F_{xy}$) are applied simultaneously, the strain triple ($\varepsilon_x$, $\varepsilon_y$, $\gamma_{xy}$) spans the entire in-plane strain space. In this case, for every admissible strain triple there exists a unique corresponding stress triple ($\sigma_x$, $\sigma_y$, $\tau_{xy}$). This is precisely the definition of a first-degree polynomial identity. The macroscopic equilibrium relations thus become polynomial identities, which requires the coefficients to match term by term. This yields a unique and load-independent set of equivalent stiffness constants.

The classical Voigt rule may be interpreted as a particular realization of this result, where the coefficient matching is postulated a priori through volume-weighted averaging. In the present formulation, however, this coefficient matching is not assumed but arises as a mathematical necessity from equilibrium, compatibility, and perfect bonding. In contrast to Voigt, where this matching is postulated, the present result follows from polynomial identity and equilibrium

3. **General Loading**: $F_x + F_y + F_{xy}$

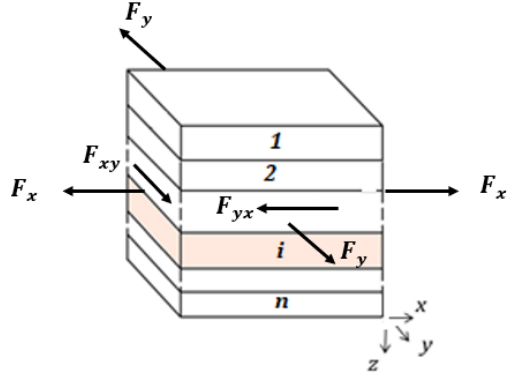

Figure 3

If all the above forces are applied simultaneously, the expressions for the resulting $\sigma_x$, $\sigma_y$ and $\tau_{xy}$ stresses will be the same on a layer and system basis.

$$\sigma_x = \sum_i V_i \, \sigma_{i_x} \tag{18}$$

$$\sigma_y = \sum_i V_i \, \sigma_{i_y} \tag{19}$$

$$\tau_{xy} = \sum_i V_i \, \tau_{i_{xy}} \tag{20}$$

Let's write the Hooke relations from the matrices (1) and (2) on a layer and system basis:

For a layer $i$

$$\sigma_{i_x} = C_{i_{11}} \varepsilon_x + C_{i_{12}} \varepsilon_y + C_{i_{16}} \gamma_{xy} \tag{21}$$

$$\sigma_{i_y} = C_{i_{12}} \varepsilon_x + C_{i_{22}} \varepsilon_y + C_{i_{26}} \gamma_{xy} \tag{22}$$

$$\tau_{i_{xy}} = C_{i_{16}} \varepsilon_x + C_{i_{26}} \varepsilon_y + C_{i_{66}} \gamma_{xy} \tag{23}$$

For all entire layered structure:

$$\sigma_x = C_{11}\varepsilon_x + C_{12}\varepsilon_y + C_{16}\gamma_{xy} \quad (24)$$

$$\sigma_y = C_{21}\varepsilon_x + C_{22}\varepsilon_y + C_{26}\gamma_{xy} \quad (25)$$

$$\tau_{xy} = C_{61}\varepsilon_x + C_{62}\varepsilon_y + C_{66}\gamma_{xy} \quad (26)$$

From Eqs. (18-26);

$$C_{11}\varepsilon_x + C_{12}\varepsilon_y + C_{16}\gamma_{xy} = \sum_i V_i\, C_{i_{11}}\varepsilon_x + \sum_i V_i\, C_{i_{12}}\varepsilon_y + \sum_i V_i\, C_{i_{16}}\gamma_{xy} \quad (27)$$

$$C_{21}\varepsilon_x + C_{22}\varepsilon_y + C_{26}\gamma_{xy} = \sum_i V_i\, C_{i_{12}}\varepsilon_x + \sum_i V_i\, C_{i_{22}}\varepsilon_y + \sum_i V_i\, C_{i_{26}}\gamma_{xy} \quad (28)$$

$$C_{61}\varepsilon_x + C_{62}\varepsilon_y + C_{66}\gamma_{xy} = \sum_i V_i\, C_{i_{16}}\varepsilon_x + \sum_i V_i\, C_{i_{26}}\varepsilon_y + \sum_i V_i\, C_{i_{66}}\gamma_{xy} \quad (29)$$

Each ($\varepsilon_x$, $\varepsilon_y$, $\gamma_{xy}$) strain triplet actually corresponds to a unique sigma value. Therefore, the above last equations must hold for all ($\varepsilon_x$, $\varepsilon_y$, $\gamma_{xy}$) triplets. This necessitates that these equations be first-degree **polynomials**. Therefore, the same strain coefficients must be equal to each other.

$$C_{11} = \sum_i V_i\, C_{i_{11}}, \quad C_{12} = \sum_i V_i\, C_{i_{12}}, \quad C_{16} = \sum_i V_i\, C_{i_{16}} \quad (30\text{a-c})$$

$$C_{21} = \sum_i V_i\, C_{i_{12}}, \quad C_{22} = \sum_i V_i\, C_{i_{22}}, \quad C_{26} = \sum_i V_i\, C_{i_{26}} \quad (30\text{d-f})$$

$$C_{61} = \sum_i V_i\, C_{i_{16}}, \quad C_{62} = \sum_i V_i\, C_{i_{26}}, \quad C_{66} = \sum_i V_i\, C_{i_{66}} \quad (30\text{g-i})$$

$$C_{12} = C_{21}, \qquad C_{16} = C_{61}, \qquad C_{26} = C_{62} \qquad (31\text{a-c})$$

As can be seen, the general loading condition only is sufficient to prove the Reciprocity conditions on a system basis.

**Conclusion**

This study has demonstrated that, for asymmetric and generally anisotropic (including triclinic) laminated structures, the elastic constants of an equivalent volume can be defined uniquely and in a load-independent manner only when the laminate is subjected to the complete set of in-plane force resultants ($F_x$, $F_y$ and $F_{xy}$). Under this general loading condition, the strain components ($\varepsilon_x$, $\varepsilon_y$, $\gamma_{xy}$) become independent and span the full in-plane strain space, which turns the macroscopic stress–strain relations into polynomial identities. As a direct mathematical consequence, the elastic constants of the equivalent volume emerge as unique volume-weighted sums of the corresponding layer constants, without any additional assumptions.

Unlike classical averaging approaches, such as the Voigt model, where the symmetry and reciprocity of the equivalent stiffness matrix are implicitly assumed, the Zor formulation shows that these properties arise automatically from equilibrium and perfect bonding between layers. Reciprocity is thus not imposed at the system level but appears as a natural outcome of the theory, consistently inherited from the layer-level constitutive behavior.

Because the resulting equivalent elastic constants are independent of the applied loading, they represent intrinsic material characteristics of the laminated structure rather than properties tied to a particular test configuration. This establishes a general and physically consistent equivalent-volume law for laminated composites, valid for arbitrary anisotropy and asymmetry, and grounded solely in linear elasticity, static equilibrium, and interlayer compatibility.

This derivation — showing that the equivalent stiffness matrix emerges uniquely from general loading through polynomial identity, rather than being postulated — constitutes the most original and fundamental contribution of the present formulation.